\documentclass[preprintnumbers,amssymb,amsmath,letterpaper,nofootinbib,floatfix]{revtex4-2}

\usepackage[dvipsnames]{xcolor}
\usepackage[colorlinks,allcolors=RoyalBlue]{hyperref}
\usepackage{graphicx}
\usepackage{bm}
\usepackage[normalem]{ulem}

\newcommand{\gsim}{\raisebox{-0.13cm}{~\shortstack{$>$ \\[-0.07cm]
      $\sim$}}~}
\newcommand{\lsim}{\raisebox{-0.13cm}{~\shortstack{$<$ \\[-0.07cm]
      $\sim$}}~}

\makeatletter
\def\bibsection{%
   \par
   \begingroup
    \baselineskip26\p@
    \bib@device{\hsize}{72\p@}%
   \endgroup
   \nobreak\@nobreaktrue
   \addvspace{19\p@}%
  }%
\makeatother

\begin{document}

\preprint{\tt FERMILAB-PUB-24-0354-T}

\title{Kaluza-Klein Graviton Freeze-In and Big Bang Nucleosynthesis}

\author{Mathieu Gross$^{1}$}
\thanks{mathieu.gross@ijclab.in2p3.fr,
https://orcid.org/0009-0009-0290-5755}

\author{Dan Hooper$^{2,3,4}$}
\thanks{dhooper@uchicago.edu, http://orcid.org/0000-0001-8837-4127}

\affiliation{$^1$Université Paris-Saclay, CNRS/IN2P3, IJCLab, 91405 Orsay, France}
\affiliation{$^2$University of Chicago, Department of Astronomy and Astrophysics, Chicago, Illinois 60637, USA}
\affiliation{$^3$University of Chicago, Kavli Institute for Cosmological Physics, Chicago, Illinois 60637, USA}
\affiliation{$^4$Fermi National Accelerator Laboratory, Theoretical Astrophysics Group, Batavia, Illinois 60510, USA}

\date{\today}

\begin{abstract}

In models featuring extra spatial dimensions, particle collisions in the early universe can produce Kaluza-Klein gravitons. Such particles will later decay, potentially impacting the process of Big Bang nucleosynthesis. In this paper, we consider scenarios in which gravity is free to propagate throughout $n$ flat, compactified extra dimensions, while the fields of the Standard Model are confined to a 3+1 dimensional brane. We calculate the production and decay rates of the states that make up the Kaluza-Klein graviton tower and determine the evolution of their abundances in the early universe. We then go on to evaluate the impact of these decays on the resulting light element abundances. We identify significant regions of previously unexplored parameter space that are inconsistent with measurements of the primordial helium and deuterium abundances. In particular, we find that for the case of one extra dimension (two extra dimensions), the fundamental scale of gravity must be $M_{\star} \gsim 2 \times 10^{13} \, {\rm GeV}$ ($M_{\star} \gsim 10^{10} \, {\rm GeV}$) unless the temperature of the early universe was never greater than $T \sim 2 \, {\rm TeV}$ ($T \sim 1 \, {\rm GeV}$). For larger values of $n$, these constraints are less stringent. For the case of $n=6$, for example, our analysis excludes all values of $M_{\star}$ less than $\sim 10^{6} \, {\rm GeV}$, unless the temperature of the universe was never greater than $T \sim 3 \, {\rm TeV}$. The results presented here severely limit the possibility that black holes were efficiently produced through particle collisions in the early universe's thermal bath. 

\end{abstract}

\maketitle

\section{Introduction}
\label{sec:intro}

Measurements of the primordial light element abundances provide us with our earliest probe of cosmic history, allowing us to constrain the expansion rate and overall composition of our universe as early as $\sim 1 \, {\rm s}$ after the Big Bang. In particular, this information indicates that the early universe was radiation dominated
and at least as hot as a few MeV~\cite{Hasegawa:2019jsa,Hannestad:2004px,Giudice:2000ex,Kawasaki:2000en,Kawasaki:1999na}. Little is known, however, about the thermal history of our universe prior to the onset of Big Bang nucleosynthesis. 

At extremely high temperatures, particles in the thermal plasma can scatter to produce gravitational excitations, leading to the production of a stochastic gravitational wave background~\cite{Ghiglieri:2020mhm,Ghiglieri:2015nfa}. Such considerations allow us to constrain the maximum temperature of the very early universe, $T_{\rm max} \lsim M_{\rm Pl} \sim 10^{19} \, {\rm GeV}$~\cite{Hu:2020wul,Ringwald:2020ist}. If those gravitational waves were later diluted, such as through inflation, this constraint would instead apply to the temperature of subsequent reheating, $T_{\rm RH} \lsim M_{\rm Pl} \sim 10^{19} \, {\rm GeV}$.

If our universe has extra spatial dimensions, gravitational excitations can be be produced more efficiently and at lower temperatures. As a result, we can potentially place much more stringent constraints on the maximum temperature of the universe at early times in such scenarios. In this study, we will focus on the model proposed by Arkani-Hamed, Dimopoulos, and Dvali (ADD)~\cite{Arkani-Hamed:1998jmv,Antoniadis:1998ig,Arkani-Hamed:1998sfv}, which features $n$ extra dimensions that are flat and compactified on a torus of radius, $R$. Unlike gravity, all of the Standard Model fields are confined to a $3$-dimensional brane, the volume of which constitutes the $3+1$ dimensional spacetime that we experience.

This class of models was originally proposed as a possible solution to the electroweak hierarchy problem~\cite{Arkani-Hamed:1998jmv}. In particular, the effective 4-dimensional Planck scale, $M_{\rm Pl} \approx 1.22 \times 10^{19} \, {\rm GeV}$, is related to the fundamental $n+4$ dimensional Planck scale, $M_{\star}$, according to the following~\cite{Arkani-Hamed:1998jmv,Arkani-Hamed:1998sfv}: 
\begin{align}
M^2_{\rm Pl} = (2\pi)^n \, R^n \, M_{\star}^{2+n}.
\end{align}
Thus, for the appropriate values of $R$ and $n$, the fundamental scale of gravity could be similar to electroweak scale, $M_{\star} \sim {\rm TeV}$. In such a scenario, the apparent hierarchy between the Planck scale and the electroweak scale would be a consequence of the Standard Model's localization on the $3+1$ dimensional brane. 

Since the ADD model was proposed more than two decades ago, stringent constraints have been placed on this class of scenarios. In particular, data from the Large Hadron Collider require that $M_{\star}$ must be greater than several TeV~\cite{CMS:2017zts,ATLAS:2017bfj}. Tests of the gravitational force law at sub-millimeter distances further constrain $R \lsim 30 \, \mu {\rm m}$~\cite{Adelberger:2009zz,Murata:2014nra,Tan:2016vwu}, corresponding to $M_{\star} \gsim 5.4\times 10^{8} \, {\rm GeV}$ for $n=1$ and $M_{\star}>3.6\times 10^{3} \, {\rm GeV}$ for $n=2$. The requirement that neutron stars are not overly heated by Kaluza-Klein graviton decays further requires $M_{\star} \gsim 1.7 \times 10^5\, {\rm GeV}$ for $n=2$ and $M_{\star}\gsim 7.6 \times 10^4\, {\rm GeV}$ for $n=3$~\cite{Hannestad:2003yd} (for further discussion, see the entry on {\it Extra Dimensions} in the Particle Data Group's Review of Particle Physics~\cite{ParticleDataGroup:2022pth}).

In light of the constraints from the Large Hadron Collider, we will not attempt to motivate this study by appealing to the electroweak hierarchy problem, but rather by the broader possibility of extra spatial dimensions, such as within the context of string theory~\cite{Antoniadis:1998ig,Shiu:1998pa}. With this in mind, we will consider values of $M_{\star}$ that range from several TeV up to the Planck scale. 

In this class of models, particle collisions in the early universe can result in the efficient production of Kaluza-Klein gravitons.  For gravitons lighter than $m_{\rm KK} \sim 10^6 \, {\rm GeV}$, such states will decay during or after the era of Big Bang Nucleosynthesis, producing energetic Standard Model particles that can break apart helium nuclei through the processes of photodissassociation or hadrodissociation. Such decays can reduce of the primordial helium abundance and increase the abundance of primordial deuterium. We explore the impact of these decays and use measurements of the light element abundances to place constraints on this class of scenarios. 

The remainder of this paper is organized as follows. In Sec.~\ref{sec:KK}, we evaluate the evolution of the Kaluza-Klein graviton abundance in the early universe, including their production via freeze-in and their subsequent decays. In Secs.~\ref{sec:BBN} and~\ref{sec:results}, we discuss the impact of these decays on the primordial light element abundances and use this information to place constraints on this class of models. In Sec.~\ref{sec:BH}, we consider the impact of these constraints on the possibility that black holes could be efficiently produced through particle collisions in the early universe. We summarize our main results in Sec.~\ref{sec:summary}.

\section{Kaluza-Klein Graviton Freeze In}
\label{sec:KK}

In the ADD scenario, all of the Standard Model fields are restricted to propagate within the $3+1$ dimensional brane. In constrast, gravitons are free to propagate throughout the $n+4$ dimensional bulk. To observers on the brane, these massless spin-$2$ gravitons appear as massive Kaluza-Klein states. More specifically, for each level of the Kaluza-Klein graviton tower, there exists one spin-$2$ state, $\tilde{h}_m$, $(n-1)$ spin-$1$ states, and $n(n-1)/2$ spin-$0$ states, $\tilde{\phi}_m$, all with masses given by $m_{\tilde{h}_m} = m_{\tilde{\phi}_m} = m/R$, where $m$ is the level of the Kaluza-Klein tower. The spin-$1$ states are entirely decoupled and will play no role in the calculations performed here~\cite{Han:1998sg}.

The spin-2 Kaluza-Klein gravitons decay to Standard Model fields with the following partial widths~\cite{Han:1998sg}:
\begin{align}
\Gamma_{\tilde{h}_m \rightarrow \gamma \gamma} &= \frac{m^3_{\tilde{h}_m}}{80\pi M^2_{\rm Pl}} \\
\Gamma_{\tilde{h}_m \rightarrow ZZ} &= \frac{13 m^3_{\tilde{h}_m}}{960 \pi M^2_{\rm Pl}}\bigg(1-\frac{4 m_Z^2}{m^2_{\tilde{h}_m}}\bigg)^{1/2} \bigg(1+\frac{56 m_Z}{169 m_{\tilde{h}_m}} + \frac{36 m_Z^2}{169 m^2_{\tilde{h}_m}}\bigg) \nonumber \\
\Gamma_{\tilde{h}_m \rightarrow WW} &= \frac{13 m^3_{\tilde{h}_m}}{480 \pi M^2_{\rm Pl}}\bigg(1-\frac{4 m_W^2}{m^2_{\tilde{h}_m}}\bigg)^{1/2} \bigg(1+\frac{56 m_W}{169 m_{\tilde{h}_m}} + \frac{36 m_W^2}{169 m^2_{\tilde{h}_m}}\bigg) \nonumber \\
\Gamma_{\tilde{h}_m \rightarrow gg} &= \frac{m^3_{\tilde{h}_m}}{10 \pi M^2_{\rm Pl}} \nonumber \\
\Gamma_{\tilde{h}_m \rightarrow HH} &= \frac{m^3_{\tilde{h}_m}}{960 \pi M^2_{\rm Pl}} \bigg(1-\frac{4 m_H^2}{m^2_{\tilde{h}_m}}\bigg)^{5/2}\nonumber \\
\Gamma_{\tilde{h}_m \rightarrow f\bar{f}} &= \frac{g_f \, m^3_{\tilde{h}_m}}{640 \pi M^2_{\rm Pl}} \bigg(1-\frac{4 m_f^2}{m^2_{\tilde{h}_m}}\bigg)^{3/2} \bigg(1+\frac{8 m_f}{3 m_{\tilde{h}_m}}\bigg), \nonumber
\end{align}
where $g_f$ is the number of internal degrees-of-freedom of the fermion species (4 for each charged lepton, 2 for each neutrino, and 12 for each quark).

%

For $m_{\tilde{h}_m} \gsim {\rm TeV}$, the sum of these partial widths is given by\footnote{Note that this quantity is sometimes calculated including decays into right-handed neutrinos, in which case the numerical prefactor is instead given by $292/960\pi=73/240\pi$.}
\begin{align}
\label{gammasum}
\Gamma_{\tilde{h}_m} \approx \frac{283 m^3_{\tilde{h}_m}}{960 \pi M^2_{\rm Pl}}. 
\end{align}
\begin{figure}[t]
\includegraphics[width=0.6\linewidth]{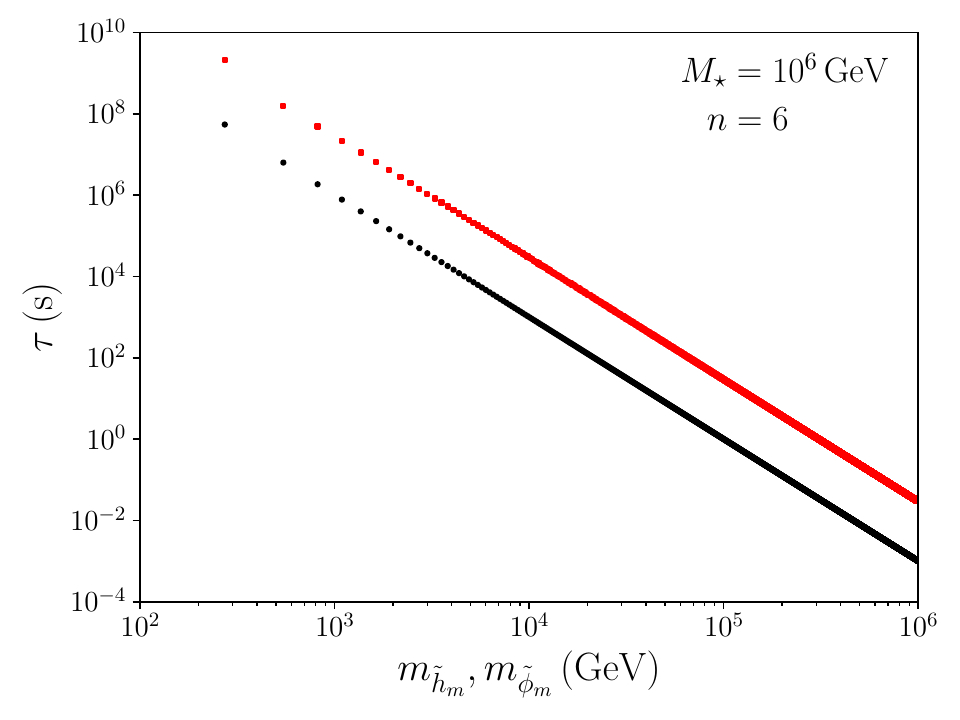} 
\caption{The lifetimes of Kaluza-Klein gravitons as a function of their mass, for the case of $M_{\star}=10^6 \, {\rm GeV}$ and $n=6$ (for which the lightest Kaluza-Klein states have a mass of $1/R \approx 273 \, {\rm GeV}$). The black dots represent the spin-2 states ($\tilde{h}_m$), while the red squares correspond to the spin-0 states ($\tilde{\phi}_m$). We expect those Kaluza-Klein modes with $\tau \gsim 1 \, {s}$ to decay after the onset of Big Bang nucleosynthesis, potentially impacting the primordial light element abundances.}
\label{fig:lifetime}
\end{figure}

The partial widths of the spin-0 Kaluza-Klein graviton states are given by~\cite{Han:1998sg}:
\begin{align}
\Gamma_{\tilde{\phi}_m \rightarrow \gamma \gamma} &= 0 \\
\Gamma_{\tilde{\phi}_m \rightarrow ZZ} &= \frac{m^3_{\tilde{\phi}_m}}{(n+2) 48 \pi M^2_{\rm Pl}}\bigg(1-\frac{4 m_Z^2}{m^2_{\tilde{\phi}_m}}\bigg)^{1/2} \bigg(1-\frac{4 m_Z}{m_{\tilde{\phi}_m}} + \frac{12 m_Z^2}{m^2_{\tilde{\phi}_m}}\bigg) \nonumber \\
\Gamma_{\tilde{\phi}_m \rightarrow WW} &= \frac{m^3_{\tilde{\phi}_m}}{(n+2) 24 \pi M^2_{\rm Pl}}\bigg(1-\frac{4 m_W^2}{m^2_{\tilde{\phi}_m}}\bigg)^{1/2} \bigg(1-\frac{4 m_W}{m_{\tilde{\phi}_m}} + \frac{12 m_W^2}{m^2_{\tilde{\phi}_m}}\bigg) \nonumber \\
\Gamma_{\tilde{\phi}_m \rightarrow gg} &= 0\nonumber \\
\Gamma_{\tilde{\phi}_m \rightarrow HH} &= \frac{m^3_{\tilde{\phi}_m}}{(n+2)48 \pi M^2_{\rm Pl}} \bigg(1-\frac{4 m_H^2}{m^2_{\tilde{\phi}_m}}\bigg)^{1/2} \bigg(1+\frac{2 m_H^2}{m^2_{\tilde{\phi}_m}}\bigg)^2 \nonumber \\
\Gamma_{\tilde{\phi}_m \rightarrow f\bar{f}} &= \frac{g_f \, m^2_f m_{\tilde{\phi}_m}}{(n+2)48 \pi  M^2_{\rm Pl}} \bigg(1-\frac{4 m_f^2}{m^2_{\tilde{\phi}_m}}\bigg)^{1/2} \bigg(1-\frac{2 m_f}{m_{\tilde{\phi}_m}}\bigg),\nonumber
\end{align}
which for $m_{\tilde{\phi}_m} \gg {\rm TeV}$ sums to
\begin{align}
\Gamma_{\tilde{\phi}_m} \approx \frac{m^3_{\tilde{\phi}_m}}{(n+2)12 \pi M^2_{\rm Pl}}. 
\end{align}
In Fig.~\ref{fig:lifetime}, we plot of the lifetimes of the Kaluza-Klein gravitons as a function of their mass, for the case of $M_{\star}=10^6 \, {\rm GeV}$ and $n=6$.

\begin{figure}[t]
\includegraphics[width=0.6\linewidth]{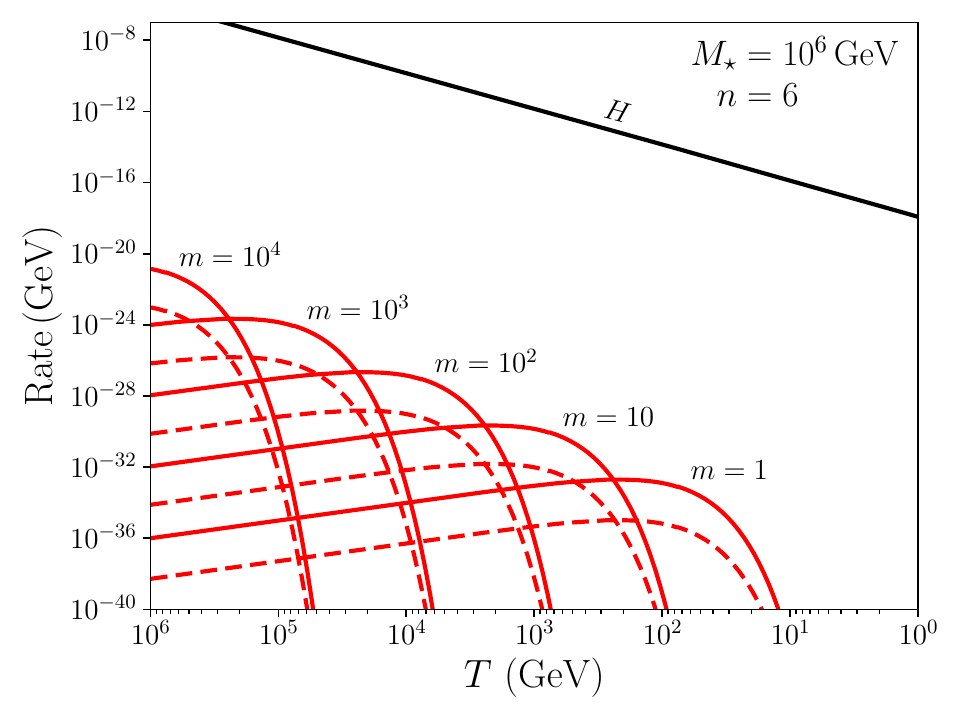} 
\caption{The production rate of Kaluza-Klein gravitons divided by three powers of temperature, $P/T^3$, for the case of $M_{\star}=10^6 \, {\rm GeV}$ and $n=6$. We show results for Kaluza-Klein gravitons with levels of $m=10^4, 10^3, 10^2, 10$, and 1. The solid red lines correspond to the spin-2 states, $\tilde{h}_m$, while the dashed red lines correspond to the scalar states, $\tilde{\phi}_m$. For comparison, we also show the value of the hubble rate, $H$, which is much larger than the production rate of Kaluza-Klein gravitons, ensuring that equilibrium is never achieved.}
\label{fig:map}
\end{figure}

Kaluza-Klein gravitons can be produced in the early universe through the inverse decays of Standard Model particles, such as $e^+ e^- \rightarrow \tilde{h}_m$, for example. It follows from the principle of detailed balance that, in equilibrium, the production rate of a given particle species will be equal to the rate of its destruction. We can use this information to calculate the production rates of the various Kaluza-Klein graviton modes from their decay widths~\cite{Hall:1999mk,deGiorgi:2021xvm,deGiorgi:2022yha}:
\begin{align}
P_{\tilde{h}_m} &= n^{\rm Eq}_{\tilde{h}_m} \, \langle \Gamma_{\tilde{h}_m} \rangle \\
&\approx n^{\rm Eq}_{\tilde{h}_m} \, \Gamma_{\tilde{h}_m}  \, \frac{K_1(m_{\tilde{h}_m}/T)}{K_2(m_{\tilde{h}_m}/T)}, \nonumber \\[3pt]
P_{\tilde{\phi}_m} &= n^{\rm Eq}_{\tilde{\phi}_m} \, \langle \Gamma_{\tilde{\phi}_m} \rangle \nonumber \\
&\approx n^{\rm Eq}_{\tilde{\phi}_m} \, \Gamma_{\tilde{\phi}_m}  \, \frac{K_1(m_{\tilde{\phi}_m}/T)}{K_2(m_{\tilde{\phi}_m}/T)},\nonumber
\end{align}
where $\langle \Gamma_{\tilde{h}_m} \rangle$ and $\langle \Gamma_{\tilde{\phi}_m} \rangle$ are the thermally averaged decay widths~\cite{Escudero:2019gzq}, $T$ is the temperature of the decaying particle population, and $K_1$ and $K_2$ are modified Bessel functions of the second kind. The equilibrium number densities of Kaluza-Klein gravitons are given by
\begin{align}
n^{\rm Eq}_{\tilde{h}_m} &= \frac{5}{2\pi^2} \int_{m_{\tilde{h}_m}}^{\infty} \frac{(E^2-m^2_{\tilde{h}_m})^{1/2}}{e^{E/T} -1} E dE \\
n^{\rm Eq}_{\tilde{\phi}_m} &= \frac{n(n-1)}{4\pi^2} \int_{m_{\tilde{\phi}_m}}^{\infty} \frac{(E^2-m^2_{\tilde{\phi}_m})^{1/2}}{e^{E/T} -1} E dE, 
\nonumber 
\end{align}
where the factor of $n(n-1)/2$ in the second expression accounts for the multiplicity of spin-0 states at each level of the Kaluza-Klein tower.

For $T \gsim {\rm TeV}$, these production rates reduce to
\begin{align}
P_{\tilde{h}_m} &\approx \frac{5 \zeta(3) T^3}{\pi^2} \, \frac{283 m^3_{\tilde{h}_m}}{960 \pi M^2_{\rm Pl}} \, \frac{K_1(m_{\tilde{h}_m}/T)}{K_2(m_{\tilde{h}_m}/T)} \\
P_{\tilde{\phi}_m} &\approx \frac{n(n-1)\zeta(3) T^3}{2\pi^2} \, \frac{m^3_{\tilde{\phi}_m}}{(n+2)12 \pi M^2_{\rm Pl}} \, \frac{K_1(m_{\tilde{\phi}_m}/T)}{K_2(m_{\tilde{\phi}_m}/T)}. \nonumber
\end{align}

Note that the production rate of spin-2 Kaluza-Klein gravitons exceeds that of spin-0 states by a factor of $\sim 283(n+2)/[16n(n-1)] \approx 4.7-35$ (for $n=6-2$). The lifetimes of the scalar modes, however, are longer than those of the spin-2 modes by a factor of $\sim 283/70$, leading to complementary impacts on the light element abundances.

We are now in a position to calculate the evolution of the abundances of the Kaluza-Klein gravitons by solving the following coupled set of differential equations (for each level of the Kaluza-Klein tower, $m$):
\begin{align}
\label{boltzmann}
\frac{dn_{\tilde{h}_m}}{dt} &= -3Hn_{\tilde{h}_m} + P_{\tilde{h}_m} - n_{\tilde{h}_m} \langle \Gamma_{\tilde{h}_m} \rangle \\
\frac{dn_{\tilde{\phi}_m}}{dt} &= -3Hn_{\tilde{\phi}_m} + P_{\tilde{\phi}_m} - n_{\tilde{\phi}_m} \langle \Gamma_{\tilde{\phi}_m} \rangle, \nonumber 
\end{align}
where $H=(8\pi \rho/3 M^2_{\rm Pl})^{1/2}$ is the rate of Hubble expansion and $\rho = \rho_{\rm SM} +\sum_m \rho_{\tilde{h}_m} + \sum_m \rho_{\tilde{\phi}_m}$ is the total energy density of the universe, $\rho_{\rm SM} = \pi^2 g_{\star}T^4/30$ is the energy density in Standard Model particles, and $g_{\star}$ is the number of relativistic degrees-of-freedom. We evolve the temperature of the Standard Model bath by applying entropy conservation, $T \propto a^{-1} g_{\star, S}^{-1/3}$, where $g_{\star, S}$ is the number of relativistic degrees-of-freedom in entropy.

In Fig.~\ref{fig:map}, we plot the production rate of Kaluza-Klein gravitons (divided by three powers of the temperature) for several selected levels of the Kaluza-Klein tower, and for the case of the $M_{\star}=10^6 \, {\rm GeV}$ and $n=6$. Since these production rates are many order of magnitude below the rate of Hubble expansion, the Kaluza-Klein graviton abundances never reach their equilibrium values, placing us safely within the regime of thermal freeze-in. Unlike more typical freeze-in scenarios, however, the large multiplicity of Kaluza-Klein states can greatly enhance the total energy density of these particles that is produced. The case shown in Fig.~\ref{fig:map}, for example, is effectively that of the simultaneous freeze-in of $\sim 2 \times 10^4$ different particle species, each of which contributes to the total resulting abundance.

\begin{figure}[t]
\includegraphics[width=0.49\linewidth]{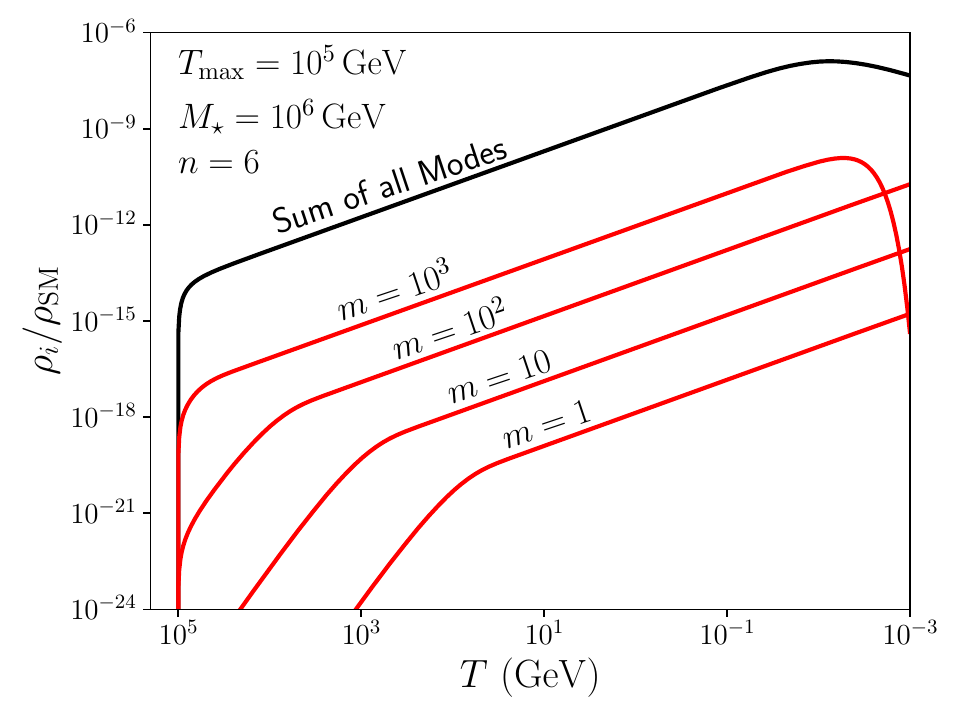} 
\includegraphics[width=0.49\linewidth]{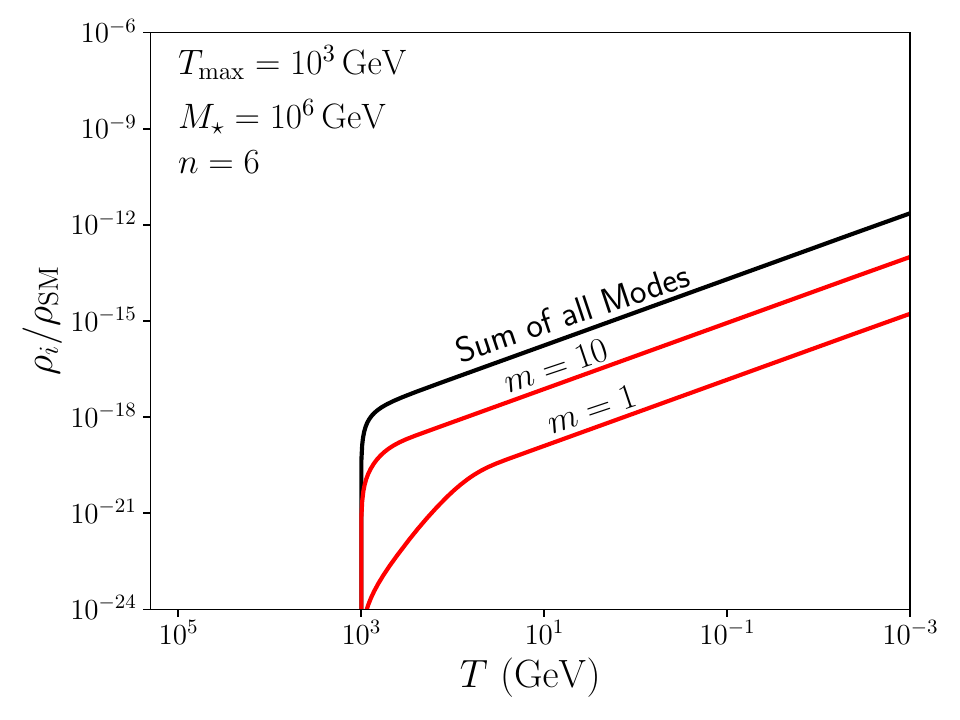} 
\caption{The accumulated non-relativistic densities of Kaluza-Klein gravitons, $\rho_i=n_i m_i$, divided by the total energy density in Standard Model particles for the case of $M_{\star}=10^6 \, {\rm GeV}$ and $n=6$, and for two values of the initial temperature, $T_{\rm max}$. We show results for the Kaluza-Klein modes with levels of $m=10^3, 10^2, 10$, and 1, as well as for the sum of all modes. The red lines represent the abundances of the spin-2 states, while the black lines include both spin-$2$ and scalar states. Note that the $m=10^3$ states begin to appreciably decay prior to the onset of Big Bang nucleosynthesis.}
\label{fig:densities}
\end{figure}

In Fig.~\ref{fig:densities}, we plot the evolution of the non-relativistic densities of Kaluza-Klein gravitons, $\rho_i=n_i m_i$, compared to the total energy density in Standard Model particles, for the case of $M_{\star}=10^6 \, {\rm GeV}$ and $n=6$, and for two values of the initial temperature, $T_{\rm max}$. Results are shown several selected values of the Kaluza-Klein levels, $m$, as well as for the sum of all modes. In the left frame, one can notice that the $m=10^3$ states begin to appreciably decay prior to the onset of Big Bang nucleosynthesis.


\section{Kaluza-Klein Graviton Decays During Big Bang Nucleosynthesis}
\label{sec:BBN}

Measurements of the primordial deuterium~\cite{Cooke:2013cba,Riemer-Sorensen:2017vxj} and helium~\cite{Izotov:2014fga,Aver:2015iza,Peimbert:2016bdg} abundances provide us with the earliest probe of our universe's thermal history, confirming that our universe was radiation dominated and generally well-described by $\Lambda$CDM cosmology throughout the era of Big Bang nucleosynthesis, which began $t \sim 1 \, {s}$ after the Big Bang~\cite{Schramm:1997vs,Steigman:2007xt,Iocco:2008va,Cyburt:2015mya,Pitrou:2018cgg,Fields:2019pfx}. Such measurements allow us to place stringent constraints on the expansion history of our universe, as well as on any energy injection that may have taken place during or after this era~\cite{Sarkar:1995dd,Pospelov:2010hj,Jedamzik:2009uy,Hufnagel:2017dgo,Keith:2020jww,Carr:2009jm,Huang:2017egl,Forestell:2018txr,Hufnagel:2018bjp,Depta:2019lbe,Kawasaki:2017bqm,Poulin:2015opa,Cyburt:2009pg,Kusakabe:2008kf,Jedamzik:2006xz,Kohri:2005wn,Jedamzik:2004er,Kawasaki:2004qu,Kawasaki:2004yh,Cyburt:2002uv,Kohri:2001jx,Kawasaki:2000qr,Holtmann:1998gd,Kawasaki:1994sc}.

\begin{figure}[t]
\includegraphics[width=0.6\linewidth]{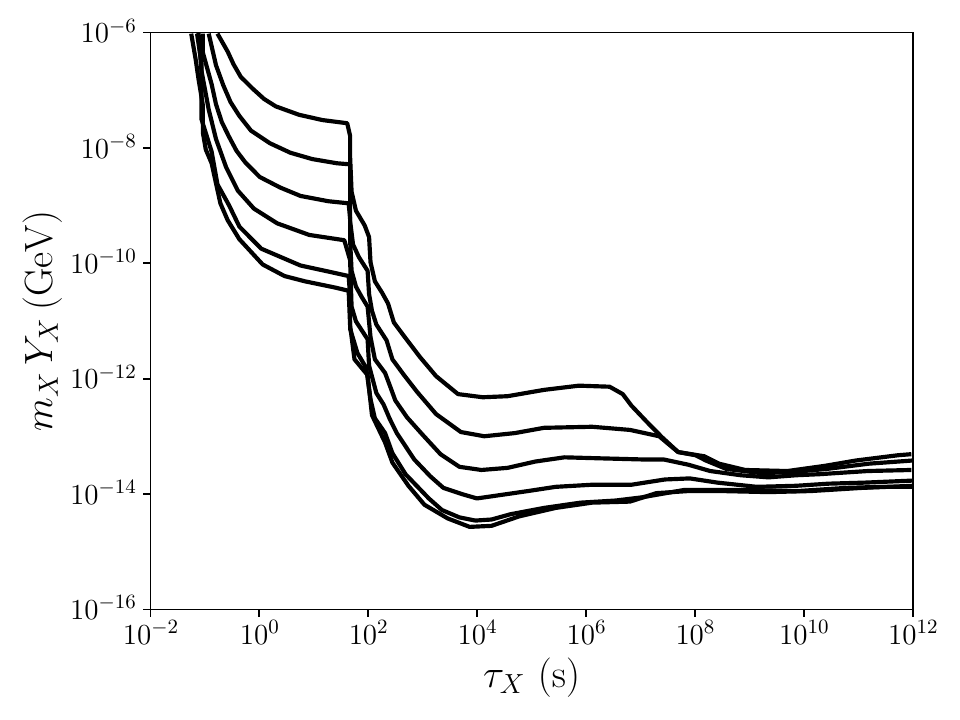} 
\caption{Constraints on a particle species decaying to quarks from measurements of the primordial element abundances, as presented in Ref.~\cite{Kawasaki:2017bqm}. From top-to-bottom, the curves correspond to the upper limits on the abundance of particles of mass $10^6 \, {\rm GeV}$, $10^5 \, {\rm GeV}$, $10^4 \, {\rm GeV}$, $10^3 \, {\rm GeV}$, $10^2 \, {\rm GeV}$, and $30 \, {\rm GeV}$. These constraints are presented in terms of the number density of decaying particles per unit entropy (prior to their decays), $Y_X = n_X/s$.}
\label{fig:BBN}
\end{figure}

The presence of Kaluza-Klein gravitons in the early universe could have potentially impacted the light element abundances in a number of ways. In particular, their decay products could have broken up helium nuclei through the processes of photodissociation and hadrodissociation, reducing the abundance of primordial helium while enhancing that of deuterium. These and other such processes have been modeled in detail in a number of publicly available codes~\cite{Arbey:2018zfh,Pitrou:2019nub,Pisanti:2007hk}. In this study, we make use of the results of Kawasaki {\it et al.}~(2018), who evaluated the impact of decaying particles on the resulting light element abundances~\cite{Kawasaki:2017bqm} (see also Refs.~\cite{Forestell:2018txr,Depta:2019lbe,Kawasaki:1994af,Poulin:2015opa,Cyburt:2009pg,Kusakabe:2008kf,Jedamzik:2006xz,Kohri:2005wn,Jedamzik:2004er,Kawasaki:2004qu,Kawasaki:2004yh,Cyburt:2002uv,Kawasaki:2000qr,Holtmann:1998gd}). In particular, the authors of Ref.~\cite{Kawasaki:2017bqm} derived constraints on the lifetime and abundance of a decaying particle species for various values of the particle's mass and dominant decay modes. These constraints were presented in terms of the mass of the decaying particles multiplied by the number of such particles per unit entropy, $MY$, as evaluated prior to their decays, $t \ll \tau$. Whereas that study considered the impact of only one decaying particle species at a time, we are concerned here with the decays of the entire tower of Kaluza-Klein gravitons. To recast the results of Kawasaki {\it et al.}~for the case at hand, we treat all Kaluza-Klein states with lifetimes within a given decade (for example, $\tau =10^2-10^3 \, {\rm s}$) as a single particle species, with a lifetime and mass equal to that of the median Kaluza-Klein state within that group. We then compare this to Fig.~12 of Kawasaki {\it et al.} in order to determine whether a given scenario is consistent with the measured helium and deuterium abundances. We repeat this procedure for each decade of lifetime, allowing us to produce conservative constraints on the values of $M_{\star}$ and $n$, as a function of the initial temperature of the universe, $T_{\rm max}$. 
We include in our calculations only those graviton decays that proceed to quarks or gluons, and apply Kawasaki's constraints on the $u\bar{u}$ channel (which are almost entirely indistinguishable from those found in the cases of other hadronic final states). The constraints are presented by Kawasaki {\it et al.} are shown in Fig.~\ref{fig:BBN}.

\begin{figure}[t]
\includegraphics[width=0.6\linewidth]{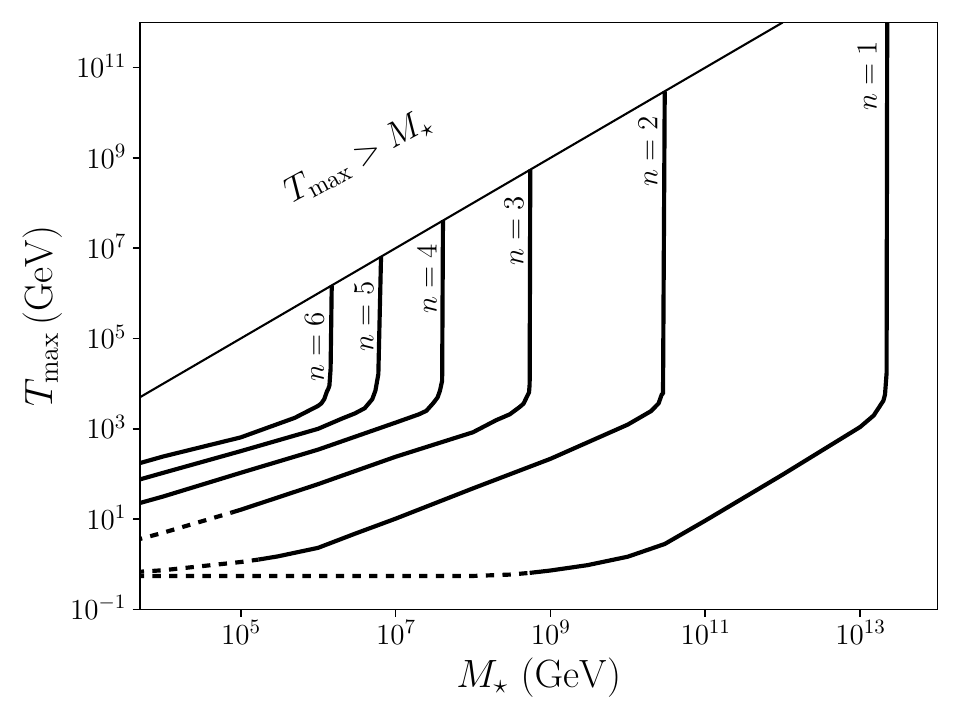} 
\caption{The upper limits derived in this study on the maximum temperature of the universe, $T_{\rm max}$, as a function of $M_{\star}$ and $n$. The dashed portions of the lines represent values of $M_{\star}$ that are ruled out by other considerations (see Sec.~\ref{sec:intro}). We only consider values of the temperatures that are below the fundamental Planck scale, $T < M_{\star}$.}
\label{fig:limit}
\end{figure}

\section{Results}
\label{sec:results}

The main results of this study are present in Fig.~\ref{fig:limit}, where we plot the maximum temperature of the universe that is consistent with the measured light element abundances, as a function of $M_{\star}$ and $n$. Where the curves are solid in this figure, the light element abundances provide the strongest constraint on this class of models. In contrast, whereas the curves are dashed, the constraints derived here are less restrictive than others that have been presented in the literature (as summarized in Sec.~\ref{sec:intro}). 

To understand the results that appear in Fig.~\ref{fig:limit}, note that we are applying constraints that apply to particles with lifetimes ranging from $\tau \sim 0.1 \, {\rm s}$ to $10^{12} \, {\rm s}$. From Eq.~\ref{gammasum}, we find that this range corresponds to Kaluza-Klein gravitons with masses in the range of $m_{\tilde{h}_m} \sim 10 \, {\rm GeV}$ to $\sim 2 \times 10^5 \, {\rm GeV}$. Thus, if the universe was never at a temperature greater than $\mathcal{O}({\rm GeV})$, the only Kaluza-Klein gravitons that could be produced would be too long-lived to be constrained by the present analysis (although other constraints, such as those from the cosmic microwave background could still be potentially restrictive). On the other hand, in order for the Kaluza-Klein gravitons to not decay prior to 0.1 s,  (or equivalently, for those states to be lighter than $m_{\tilde{h}_m} \sim 2 \times 10^5 \, {\rm GeV}$), the value of $M_{\star}$ must not be too high. For the $n=1$ case, for example, the first Kaluza-Klein mode has a mass of $m_{\tilde{h}_1} \sim 2 \times 10^5 \, {\rm GeV}$ for $M_{\star} \sim 10^{14} \, {\rm GeV}$, explaining why our constraints do not extend to larger values of the fundamental Planck scale. Alternatively, for $n=6$, $m_{\tilde{h}_1} \sim 2 \times 10^5 \, {\rm GeV}$ corresponds to a value of $M_{\star} \sim 10^{8} \, {\rm GeV}$. In this later case, you may notice that our constraints only extend up to $M_{\star} \sim 10^{6} \, {\rm GeV}$. The reason that they do not extend to higher values of $M_{\star}$ is that the number of decaying particles per unit entropy is not particular large in this case and the decays occur around $\tau \lsim 0.1 \, {\rm s}$, where the constraints are rather weak (see Fig.~\ref{fig:BBN}). If we instead consider $M_{\star} \sim 10^{6} \, {\rm GeV}$, the lightest Kaluza-Klein mode has a mass of $m_{\tilde{h}_1} \sim 300 \, {\rm GeV}$ and a lifetime of $\tau \sim 4 \times 10^7 \, {\rm s}$. Constraints in this lifetime range are much more stringent, allowing us to exclude values of $T_{\rm max}$ that are comparable to or larger than the mass of the lightest Kaluza-Klein mode.


\section{Implications for Black Hole Production}
\label{sec:BH}

It has long been appreciated that small-scale inhomogeneties in the early universe may have led to the formation of primordial
black holes~\cite{Carr:1974nx,Carr:1975qj}. Alternatively, in models with extra spatial dimensions, black holes could have been produced through the collisions of particles in the thermal bath, in particular if the total energy of a collision exceeds the fundamental Planck scale, $M_{\star}$~\cite{Conley:2006jg} (see also, Refs.~\cite{Argyres:1998qn,Myers:1986un}).

\begin{figure}[t]
\includegraphics[width=0.6\linewidth]{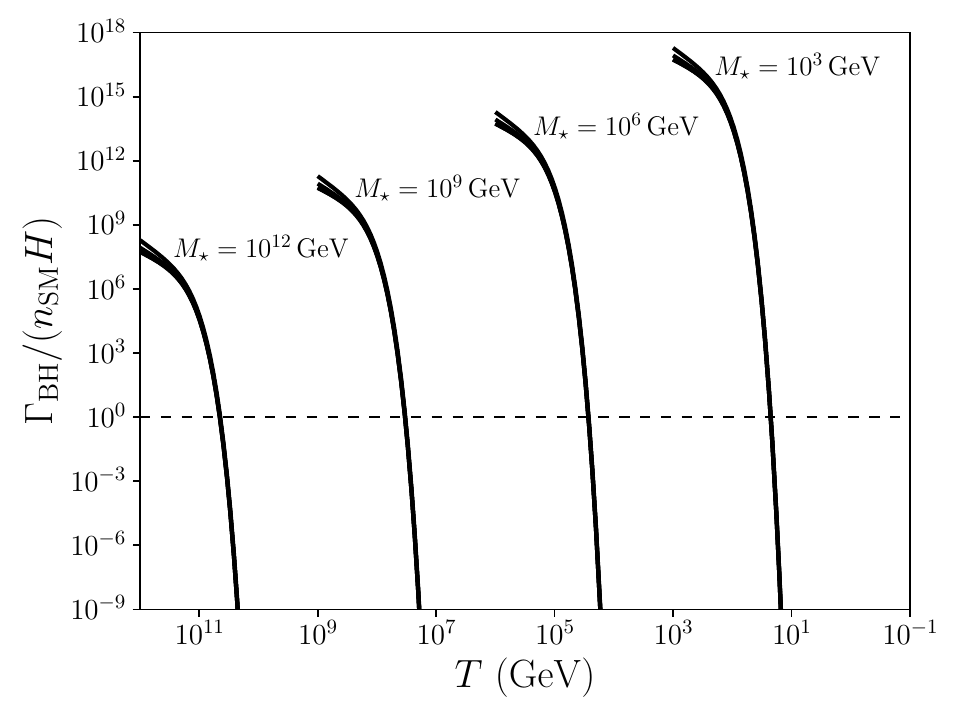} 
\caption{The integrated production rate of black holes per Standard Model particle per Hubble time, for several values of $M_{\star}$ and for $n=1,2,4$ and 6 (this result depends only very weakly on the value of $n$). For comparison, we plot as a dashed curve the value at which each Standard Model particles produces an average of one black hole per Hubble time, $\Gamma_{\rm BH}/(n_{\rm SM}H) =1$.}
\label{fig:BH}
\end{figure}

In the context of the flat and compactified extra dimensions that we have considered in this study, the differential production rate of black holes in a thermal bath of temperature, $T$, is given by~\cite{Friedlander:2022ttk}
\begin{align}
\frac{d\Gamma_{\rm BH}}{dM_{\rm BH}} = \frac{g_{\star}(T)^2}{8 \pi^4} 
\bigg(\frac{8 \Gamma(\frac{n+3}{2})}{n+2}\bigg)^{\frac{2}{n+1}}
M_{\rm BH} \, T^2 \, \bigg(\frac{M_{\rm BH}}{M_{\star}}\bigg)^{\frac{2n+4}{n+1}} 
\bigg[\frac{M_{\rm BH}}{T} K_1(M_{\rm BH}/T) + 2 K_2(M_{\rm BH}/T)\bigg] \, \Theta(M_{\rm BH}-M_{\star}),
\end{align}
where $\Gamma$ is the gamma function, $K_1$ and $K_2$ are again the modified Bessel functions of the second kind, and $\Theta$ is the Heavyside step function. For $T \sim M_{\star}$, black holes can be produced at a very high rate. In particular, after dropping order one factors, this expression integrates to $\Gamma_{\rm BH}\sim M_{\star}^4$ for temperatures near the fundamental Planck scale. This indicates that the black hole production rate per particle could potentially be larger than the Hubble rate by a huge factor, roughly $\sim (\Gamma_{\rm BH}/n_{\rm SM})/H \sim 4(M_{\star}^4/T^3)/(T^2/M_{\rm Pl}) \sim 4 M_{\rm Pl}/M_{\star}$. In contrast, the black hole production rate is dramatically suppressed at temperatures lower than $M_{\star}$. This behavior is confirmed in Fig~\ref{fig:BH}, where we plot the integrated production rate of black holes per Standard Model particle per Hubble time, for several values of $M_{\star}$ and for $n=1,2,4$ and 6 (this result depends only very weakly on the value of $n$).

If a large number of microscopic black holes had been generated in the early universe, these objects could have had a number of potentially observable impacts. In particular, the products of their Hawking evaporation could have included particles that would act as dark radiation (and contribute to the effective number of neutrino species, $N_{\rm eff}$), or contribute to the dark matter density~\cite{Friedlander:2023qmc} (see also, Refs.~\cite{Hooper:2019gtx,Hooper:2020evu,Friedlander:2023jzw}). Alternatively, in some scenarios microscopic black holes could grow rapidly through accretion, significantly delaying their evaporation~\cite{Conley:2006jg,Guedens:2002sd,Majumdar:2002mra}. 

The constraints obtained in this study severely limit the rate at which black holes could have been produced through particle collisions in the early universe. This can be seen by comparing the black hole production rates shown in Fig.~\ref{fig:BH} to the constraints we have presented in Fig.~\ref{fig:limit}. For the case of $M_{\star}=10^6 \, {\rm GeV}$, for example, our constraints derived from the primoridal light element abundances require $T_{\rm max} \lsim 3.2 \, {\rm TeV}$ for any value of $n \le 6$. For such temperatures, the black hole production rate is vanishingly small, $\Gamma_{\rm BH}/(n_{\rm SM} H) \lsim 10^{-120}$. Thus, for this value of $M_{\star}$, particle collisions will not produce any significant abundance of black holes in the early universe. If we consider larger values of $M_{\star}$, black hole production could still be potentially important. For $M_{\star}=10^9 \, {\rm GeV}$ ($10^{12} \, {\rm GeV}$), for example, large black hole production rates are still possible, provided that $n \ge 3$ ($n\ge 2$).

\section{Summary and Discussion}
\label{sec:summary}

In this study, we have considered the impact of Kaluza-Klein gravitons on the primordial light element abundances, as established during the era of Big Bang nucleosynthesis. In particular, we have focused on the ADD-scenario, which features $n=1-6$ extra spatial dimensions which are flat and compactified around a torus of radius, $R$. We have taken the fields of the Standard Model to be confined to a 3+1 dimensional brane, allowing the fundamental scale of gravity to be much lower than the effective 4-dimensional Planck scale, $M_{\star} \ll M_{\rm Pl}$~\cite{Arkani-Hamed:1998jmv,Arkani-Hamed:1998jmv,Arkani-Hamed:1998sfv}.

In this model, massless spin-2 gravitons propagating in the $n+4$ dimensional bulk appear to observers on the brane as a tower of massive Kaluza-Klein states, with an evenly-spaced series of masses, $m_{\rm KK} = m/R$, where $m$ is the level of the Kaluza-Klein tower. These Kaluza-Klein gravitons can be produced in the early universe and, for $m_{\rm KK} \lsim 10^6 \, {\rm GeV}$, will subsequently decay during or after the era of Big Bang nucleosynthesis. Such decays can produce energetic Standard Model particles which can break up helium nuclei, significantly altering the predicted abundances of primoridal helium and deuterium. 

We have calculated the production and decay rates of Kaluza-Klein gravitons in the early universe, and evaluated the impact of these decays on the primordial light element abundances. These abundances, in turn, allow us to place constraints on Kalzua-Klein gravitons with masses in the range of $m_{\rm KK} \sim 10 \, {\rm GeV}$ to $\sim 2\times 10^5 \, {\rm GeV}$, as summarized in Fig.~\ref{fig:limit}. For one extra dimension, $n=1$, we exclude all values of $M_{\star} \lsim 2 \times 10^{13} \, {\rm GeV}$ ($\lsim 10^{10} \, {\rm GeV}$) unless the maximum temperature of the universe was less than $\sim 2 \, {\rm TeV}$ ($\sim 1 \, {\rm GeV}$). For larger values of $n$, our constraints are somewhat less stringent. For the case of $n=6$, for example, our analysis excludes all values of $M_{\star} \lsim (1-2) \times 10^{6} \, {\rm GeV}$ unless the maximum temperature of the universe was less than $\sim 3 \, {\rm TeV}$. 

The results presented here severely limit the possibility that black holes may have been produced in significant numbers through particle collisions in the early universe's thermal bath~\cite{Friedlander:2022ttk}. For the case of $M_{\star} \lsim 10^6 \, {\rm GeV}$, we can rule out the possibility that any appreciable abundance of black holes was formed through such collisions, for any value of $n \le 6$. For larger values of $M_{\star}$, thermal black hole production may still have been potentially important. For $M_{\star}=10^9 \, {\rm GeV}$, for example, efficient black hole production could have occurred in the early universe, provided that $n \ge 3$.

Before closing, we note that the models we are considering here are similar to, but not the same as, the ``Dark Dimension'' scenario that has been proposed within the context of the Swampland program of string theory~\cite{Gonzalo:2022jac,Montero:2022prj}. In particular, whereas Kaluza-Klein gravitons can decay into lighter KK modes in the Dark Dimension picture, extra-dimensional momentum conservation prevents such decays in the ADD model (decays into Standard Model particles are allowed, as the brane is a dynamical object which can recoil to take on the momentum of the decaying graviton). We leave the consideration of this class of models to future work.

\vspace{16pt}

\textbf{Acknowledgments} 
\vspace{8pt}

We would like to thank Gordan Krnjaic, Yann Mambrini, Jamie Law-Smith, and Joe Lykken for helpful discussions. DH is supported by the Fermi Research Alliance, LLC under Contract No.~DE-AC02-07CH11359 with the U.S. Department of Energy, Office of High Energy Physics. This project has also received funding from the European Union’s Horizon Europe research and innovation program under the Marie Skłodowska-Curie Staff Exchange grant agreement No.~101086085.

\bibliographystyle{utphys}
\bibliography{bib2023}

\providecommand{\href}[2]{#2}\begingroup\raggedright\begin{thebibliography}{10}

\bibitem{Hasegawa:2019jsa}
T.~Hasegawa, N.~Hiroshima, K.~Kohri, R.~S.~L. Hansen, T.~Tram, and S.~Hannestad, ``{MeV-scale reheating temperature and thermalization of oscillating neutrinos by radiative and hadronic decays of massive particles},'' \href{https://dx.doi.org/10.1088/1475-7516/2019/12/012}{{\em JCAP} {\bfseries 12} (2019) 012}, \href{https://arxiv.org/abs/1908.10189}{{\ttfamily arXiv:1908.10189 [hep-ph]}}.

\bibitem{Hannestad:2004px}
S.~Hannestad, ``{What is the lowest possible reheating temperature?},'' \href{https://dx.doi.org/10.1103/PhysRevD.70.043506}{{\em Phys. Rev. D} {\bfseries 70} (2004) 043506}, \href{https://arxiv.org/abs/astro-ph/0403291}{{\ttfamily arXiv:astro-ph/0403291}}.

\bibitem{Giudice:2000ex}
G.~F. Giudice, E.~W. Kolb, and A.~Riotto, ``{Largest temperature of the radiation era and its cosmological implications},'' \href{https://dx.doi.org/10.1103/PhysRevD.64.023508}{{\em Phys. Rev. D} {\bfseries 64} (2001) 023508}, \href{https://arxiv.org/abs/hep-ph/0005123}{{\ttfamily arXiv:hep-ph/0005123}}.

\bibitem{Kawasaki:2000en}
M.~Kawasaki, K.~Kohri, and N.~Sugiyama, ``{MeV scale reheating temperature and thermalization of neutrino background},'' \href{https://dx.doi.org/10.1103/PhysRevD.62.023506}{{\em Phys. Rev. D} {\bfseries 62} (2000) 023506}, \href{https://arxiv.org/abs/astro-ph/0002127}{{\ttfamily arXiv:astro-ph/0002127}}.

\bibitem{Kawasaki:1999na}
M.~Kawasaki, K.~Kohri, and N.~Sugiyama, ``{Cosmological constraints on late time entropy production},'' \href{https://dx.doi.org/10.1103/PhysRevLett.82.4168}{{\em Phys. Rev. Lett.} {\bfseries 82} (1999) 4168}, \href{https://arxiv.org/abs/astro-ph/9811437}{{\ttfamily arXiv:astro-ph/9811437}}.

\bibitem{Ghiglieri:2020mhm}
J.~Ghiglieri, G.~Jackson, M.~Laine, and Y.~Zhu, ``{Gravitational wave background from Standard Model physics: Complete leading order},'' \href{https://dx.doi.org/10.1007/JHEP07(2020)092}{{\em JHEP} {\bfseries 07} (2020) 092}, \href{https://arxiv.org/abs/2004.11392}{{\ttfamily arXiv:2004.11392 [hep-ph]}}.

\bibitem{Ghiglieri:2015nfa}
J.~Ghiglieri and M.~Laine, ``{Gravitational wave background from Standard Model physics: Qualitative features},'' \href{https://dx.doi.org/10.1088/1475-7516/2015/07/022}{{\em JCAP} {\bfseries 07} (2015) 022}, \href{https://arxiv.org/abs/1504.02569}{{\ttfamily arXiv:1504.02569 [hep-ph]}}.

\bibitem{Hu:2020wul}
B.~X. Hu and A.~Loeb, ``{An Upper Limit on the Initial Temperature of the Radiation-Dominated Universe},'' \href{https://dx.doi.org/10.1088/1475-7516/2021/01/041}{{\em JCAP} {\bfseries 01} (2021) 041}, \href{https://arxiv.org/abs/2004.02895}{{\ttfamily arXiv:2004.02895 [astro-ph.CO]}}.

\bibitem{Ringwald:2020ist}
A.~Ringwald, J.~Sch\"utte-Engel, and C.~Tamarit, ``{Gravitational Waves as a Big Bang Thermometer},'' \href{https://dx.doi.org/10.1088/1475-7516/2021/03/054}{{\em JCAP} {\bfseries 03} (2021) 054}, \href{https://arxiv.org/abs/2011.04731}{{\ttfamily arXiv:2011.04731 [hep-ph]}}.

\bibitem{Arkani-Hamed:1998jmv}
N.~Arkani-Hamed, S.~Dimopoulos, and G.~R. Dvali, ``{The Hierarchy problem and new dimensions at a millimeter},'' \href{https://dx.doi.org/10.1016/S0370-2693(98)00466-3}{{\em Phys. Lett. B} {\bfseries 429} (1998) 263--272}, \href{https://arxiv.org/abs/hep-ph/9803315}{{\ttfamily arXiv:hep-ph/9803315}}.

\bibitem{Antoniadis:1998ig}
I.~Antoniadis, N.~Arkani-Hamed, S.~Dimopoulos, and G.~R. Dvali, ``{New dimensions at a millimeter to a Fermi and superstrings at a TeV},'' \href{https://dx.doi.org/10.1016/S0370-2693(98)00860-0}{{\em Phys. Lett. B} {\bfseries 436} (1998) 257--263}, \href{https://arxiv.org/abs/hep-ph/9804398}{{\ttfamily arXiv:hep-ph/9804398}}.

\bibitem{Arkani-Hamed:1998sfv}
N.~Arkani-Hamed, S.~Dimopoulos, and G.~R. Dvali, ``{Phenomenology, astrophysics and cosmology of theories with submillimeter dimensions and TeV scale quantum gravity},'' \href{https://dx.doi.org/10.1103/PhysRevD.59.086004}{{\em Phys. Rev. D} {\bfseries 59} (1999) 086004}, \href{https://arxiv.org/abs/hep-ph/9807344}{{\ttfamily arXiv:hep-ph/9807344}}.

\bibitem{CMS:2017zts}
{\bfseries CMS} Collaboration, A.~M. Sirunyan {\em et~al.}, ``{Search for new physics in final states with an energetic jet or a hadronically decaying $W$ or $Z$ boson and transverse momentum imbalance at $\sqrt{s}=13\text{ }\text{ }\mathrm{TeV}$},'' \href{https://dx.doi.org/10.1103/PhysRevD.97.092005}{{\em Phys. Rev. D} {\bfseries 97} no.~9, (2018) 092005}, \href{https://arxiv.org/abs/1712.02345}{{\ttfamily arXiv:1712.02345 [hep-ex]}}.

\bibitem{ATLAS:2017bfj}
{\bfseries ATLAS} Collaboration, M.~Aaboud {\em et~al.}, ``{Search for dark matter and other new phenomena in events with an energetic jet and large missing transverse momentum using the ATLAS detector},'' \href{https://dx.doi.org/10.1007/JHEP01(2018)126}{{\em JHEP} {\bfseries 01} (2018) 126}, \href{https://arxiv.org/abs/1711.03301}{{\ttfamily arXiv:1711.03301 [hep-ex]}}.

\bibitem{Adelberger:2009zz}
E.~G. Adelberger, J.~H. Gundlach, B.~R. Heckel, S.~Hoedl, and S.~Schlamminger, ``{Torsion balance experiments: A low-energy frontier of particle physics},'' \href{https://dx.doi.org/10.1016/j.ppnp.2008.08.002}{{\em Prog. Part. Nucl. Phys.} {\bfseries 62} (2009) 102--134}.

\bibitem{Murata:2014nra}
J.~Murata and S.~Tanaka, ``{A review of short-range gravity experiments in the LHC era},'' \href{https://dx.doi.org/10.1088/0264-9381/32/3/033001}{{\em Class. Quant. Grav.} {\bfseries 32} no.~3, (2015) 033001}, \href{https://arxiv.org/abs/1408.3588}{{\ttfamily arXiv:1408.3588 [hep-ex]}}.

\bibitem{Tan:2016vwu}
W.-H. Tan, S.-Q. Yang, C.-G. Shao, J.~Li, A.-B. Du, B.-F. Zhan, Q.-L. Wang, P.-S. Luo, L.-C. Tu, and J.~Luo, ``{New Test of the Gravitational Inverse-Square Law at the Submillimeter Range with Dual Modulation and Compensation},'' \href{https://dx.doi.org/10.1103/PhysRevLett.116.131101}{{\em Phys. Rev. Lett.} {\bfseries 116} no.~13, (2016) 131101}.

\bibitem{Hannestad:2003yd}
S.~Hannestad and G.~G. Raffelt, ``{Supernova and neutron star limits on large extra dimensions reexamined},'' \href{https://dx.doi.org/10.1103/PhysRevD.69.029901}{{\em Phys. Rev. D} {\bfseries 67} (2003) 125008}, \href{https://arxiv.org/abs/hep-ph/0304029}{{\ttfamily arXiv:hep-ph/0304029}}. [Erratum: Phys.Rev.D 69, 029901 (2004)].

\bibitem{ParticleDataGroup:2022pth}
{\bfseries Particle Data Group} Collaboration, R.~L. Workman {\em et~al.}, ``{Review of Particle Physics},'' \href{https://dx.doi.org/10.1093/ptep/ptac097}{{\em PTEP} {\bfseries 2022} (2022) 083C01}.

\bibitem{Shiu:1998pa}
G.~Shiu and S.~H.~H. Tye, ``{TeV scale superstring and extra dimensions},'' \href{https://dx.doi.org/10.1103/PhysRevD.58.106007}{{\em Phys. Rev. D} {\bfseries 58} (1998) 106007}, \href{https://arxiv.org/abs/hep-th/9805157}{{\ttfamily arXiv:hep-th/9805157}}.

\bibitem{Han:1998sg}
T.~Han, J.~D. Lykken, and R.-J. Zhang, ``{On Kaluza-Klein states from large extra dimensions},'' \href{https://dx.doi.org/10.1103/PhysRevD.59.105006}{{\em Phys. Rev. D} {\bfseries 59} (1999) 105006}, \href{https://arxiv.org/abs/hep-ph/9811350}{{\ttfamily arXiv:hep-ph/9811350}}.

\bibitem{Hall:1999mk}
L.~J. Hall and D.~Tucker-Smith, ``{Cosmological constraints on theories with large extra dimensions},'' \href{https://dx.doi.org/10.1103/PhysRevD.60.085008}{{\em Phys. Rev. D} {\bfseries 60} (1999) 085008}, \href{https://arxiv.org/abs/hep-ph/9904267}{{\ttfamily arXiv:hep-ph/9904267}}.

\bibitem{deGiorgi:2021xvm}
A.~de~Giorgi and S.~Vogl, ``{Dark matter interacting via a massive spin-2 mediator in warped extra-dimensions},'' \href{https://dx.doi.org/10.1007/JHEP11(2021)036}{{\em JHEP} {\bfseries 11} (2021) 036}, \href{https://arxiv.org/abs/2105.06794}{{\ttfamily arXiv:2105.06794 [hep-ph]}}.

\bibitem{deGiorgi:2022yha}
A.~de~Giorgi and S.~Vogl, ``{Warm dark matter from a gravitational freeze-in in extra dimensions},'' \href{https://dx.doi.org/10.1007/JHEP04(2023)032}{{\em JHEP} {\bfseries 04} (2023) 032}, \href{https://arxiv.org/abs/2208.03153}{{\ttfamily arXiv:2208.03153 [hep-ph]}}.

\bibitem{Escudero:2019gzq}
M.~Escudero, D.~Hooper, G.~Krnjaic, and M.~Pierre, ``{Cosmology with A Very Light L$_{\mu}$ \ensuremath{-} L$_{\tau}$ Gauge Boson},'' \href{https://dx.doi.org/10.1007/JHEP03(2019)071}{{\em JHEP} {\bfseries 03} (2019) 071}, \href{https://arxiv.org/abs/1901.02010}{{\ttfamily arXiv:1901.02010 [hep-ph]}}.

\bibitem{Cooke:2013cba}
R.~Cooke, M.~Pettini, R.~A. Jorgenson, M.~T. Murphy, and C.~C. Steidel, ``{Precision measures of the primordial abundance of deuterium},'' \href{https://dx.doi.org/10.1088/0004-637X/781/1/31}{{\em Astrophys. J.} {\bfseries 781} no.~1, (2014) 31}, \href{https://arxiv.org/abs/1308.3240}{{\ttfamily arXiv:1308.3240 [astro-ph.CO]}}.

\bibitem{Riemer-Sorensen:2017vxj}
S.~Riemer-S\o{}rensen and E.~S. Jenssen, ``{Nucleosynthesis Predictions and High-Precision Deuterium Measurements},'' \href{https://dx.doi.org/10.3390/universe3020044}{{\em Universe} {\bfseries 3} no.~2, (2017) 44}, \href{https://arxiv.org/abs/1705.03653}{{\ttfamily arXiv:1705.03653 [astro-ph.CO]}}.

\bibitem{Izotov:2014fga}
Y.~I. Izotov, T.~X. Thuan, and N.~G. Guseva, ``{A new determination of the primordial He abundance using the He i $\lambda$10830 \r{A} emission line: cosmological implications},'' \href{https://dx.doi.org/10.1093/mnras/stu1771}{{\em Mon. Not. Roy. Astron. Soc.} {\bfseries 445} no.~1, (2014) 778--793}, \href{https://arxiv.org/abs/1408.6953}{{\ttfamily arXiv:1408.6953 [astro-ph.CO]}}.

\bibitem{Aver:2015iza}
E.~Aver, K.~A. Olive, and E.~D. Skillman, ``{The effects of He I \ensuremath{\lambda}10830 on helium abundance determinations},'' \href{https://dx.doi.org/10.1088/1475-7516/2015/07/011}{{\em JCAP} {\bfseries 07} (2015) 011}, \href{https://arxiv.org/abs/1503.08146}{{\ttfamily arXiv:1503.08146 [astro-ph.CO]}}.

\bibitem{Peimbert:2016bdg}
A.~Peimbert, M.~Peimbert, and V.~Luridiana, ``{The primordial helium abundance and the number of neutrino families},'' {\em Rev. Mex. Astron. Astrofis.} {\bfseries 52} no.~2, (2016) 419--424, \href{https://arxiv.org/abs/1608.02062}{{\ttfamily arXiv:1608.02062 [astro-ph.CO]}}.

\bibitem{Schramm:1997vs}
D.~N. Schramm and M.~S. Turner, ``{Big Bang Nucleosynthesis Enters the Precision Era},'' \href{https://dx.doi.org/10.1103/RevModPhys.70.303}{{\em Rev. Mod. Phys.} {\bfseries 70} (1998) 303--318}, \href{https://arxiv.org/abs/astro-ph/9706069}{{\ttfamily arXiv:astro-ph/9706069}}.

\bibitem{Steigman:2007xt}
G.~Steigman, ``{Primordial Nucleosynthesis in the Precision Cosmology Era},'' \href{https://dx.doi.org/10.1146/annurev.nucl.56.080805.140437}{{\em Ann. Rev. Nucl. Part. Sci.} {\bfseries 57} (2007) 463--491}, \href{https://arxiv.org/abs/0712.1100}{{\ttfamily arXiv:0712.1100 [astro-ph]}}.

\bibitem{Iocco:2008va}
F.~Iocco, G.~Mangano, G.~Miele, O.~Pisanti, and P.~D. Serpico, ``{Primordial Nucleosynthesis: from precision cosmology to fundamental physics},'' \href{https://dx.doi.org/10.1016/j.physrep.2009.02.002}{{\em Phys. Rept.} {\bfseries 472} (2009) 1--76}, \href{https://arxiv.org/abs/0809.0631}{{\ttfamily arXiv:0809.0631 [astro-ph]}}.

\bibitem{Cyburt:2015mya}
R.~H. Cyburt, B.~D. Fields, K.~A. Olive, and T.-H. Yeh, ``{Big Bang Nucleosynthesis: 2015},'' \href{https://dx.doi.org/10.1103/RevModPhys.88.015004}{{\em Rev. Mod. Phys.} {\bfseries 88} (2016) 015004}, \href{https://arxiv.org/abs/1505.01076}{{\ttfamily arXiv:1505.01076 [astro-ph.CO]}}.

\bibitem{Pitrou:2018cgg}
C.~Pitrou, A.~Coc, J.-P. Uzan, and E.~Vangioni, ``{Precision big bang nucleosynthesis with improved Helium-4 predictions},'' \href{https://dx.doi.org/10.1016/j.physrep.2018.04.005}{{\em Phys. Rept.} {\bfseries 754} (2018) 1--66}, \href{https://arxiv.org/abs/1801.08023}{{\ttfamily arXiv:1801.08023 [astro-ph.CO]}}.

\bibitem{Fields:2019pfx}
B.~D. Fields, K.~A. Olive, T.-H. Yeh, and C.~Young, ``{Big-Bang Nucleosynthesis after Planck},'' \href{https://dx.doi.org/10.1088/1475-7516/2020/03/010}{{\em JCAP} {\bfseries 03} (2020) 010}, \href{https://arxiv.org/abs/1912.01132}{{\ttfamily arXiv:1912.01132 [astro-ph.CO]}}. [Erratum: JCAP 11, E02 (2020)].

\bibitem{Sarkar:1995dd}
S.~Sarkar, ``{Big bang nucleosynthesis and physics beyond the standard model},'' \href{https://dx.doi.org/10.1088/0034-4885/59/12/001}{{\em Rept. Prog. Phys.} {\bfseries 59} (1996) 1493--1610}, \href{https://arxiv.org/abs/hep-ph/9602260}{{\ttfamily arXiv:hep-ph/9602260}}.

\bibitem{Pospelov:2010hj}
M.~Pospelov and J.~Pradler, ``{Big Bang Nucleosynthesis as a Probe of New Physics},'' \href{https://dx.doi.org/10.1146/annurev.nucl.012809.104521}{{\em Ann. Rev. Nucl. Part. Sci.} {\bfseries 60} (2010) 539--568}, \href{https://arxiv.org/abs/1011.1054}{{\ttfamily arXiv:1011.1054 [hep-ph]}}.

\bibitem{Jedamzik:2009uy}
K.~Jedamzik and M.~Pospelov, ``{Big Bang Nucleosynthesis and Particle Dark Matter},'' \href{https://dx.doi.org/10.1088/1367-2630/11/10/105028}{{\em New J. Phys.} {\bfseries 11} (2009) 105028}, \href{https://arxiv.org/abs/0906.2087}{{\ttfamily arXiv:0906.2087 [hep-ph]}}.

\bibitem{Hufnagel:2017dgo}
M.~Hufnagel, K.~Schmidt-Hoberg, and S.~Wild, ``{BBN constraints on MeV-scale dark sectors. Part I. Sterile decays},'' \href{https://dx.doi.org/10.1088/1475-7516/2018/02/044}{{\em JCAP} {\bfseries 02} (2018) 044}, \href{https://arxiv.org/abs/1712.03972}{{\ttfamily arXiv:1712.03972 [hep-ph]}}.

\bibitem{Keith:2020jww}
C.~Keith, D.~Hooper, N.~Blinov, and S.~D. McDermott, ``{Constraints on Primordial Black Holes From Big Bang Nucleosynthesis Revisited},'' \href{https://dx.doi.org/10.1103/PhysRevD.102.103512}{{\em Phys. Rev. D} {\bfseries 102} no.~10, (2020) 103512}, \href{https://arxiv.org/abs/2006.03608}{{\ttfamily arXiv:2006.03608 [astro-ph.CO]}}.

\bibitem{Carr:2009jm}
B.~J. Carr, K.~Kohri, Y.~Sendouda, and J.~Yokoyama, ``{New cosmological constraints on primordial black holes},'' \href{https://dx.doi.org/10.1103/PhysRevD.81.104019}{{\em Phys. Rev. D} {\bfseries 81} (2010) 104019}, \href{https://arxiv.org/abs/0912.5297}{{\ttfamily arXiv:0912.5297 [astro-ph.CO]}}.

\bibitem{Huang:2017egl}
G.-y. Huang, T.~Ohlsson, and S.~Zhou, ``{Observational Constraints on Secret Neutrino Interactions from Big Bang Nucleosynthesis},'' \href{https://dx.doi.org/10.1103/PhysRevD.97.075009}{{\em Phys. Rev. D} {\bfseries 97} no.~7, (2018) 075009}, \href{https://arxiv.org/abs/1712.04792}{{\ttfamily arXiv:1712.04792 [hep-ph]}}.

\bibitem{Forestell:2018txr}
L.~Forestell, D.~E. Morrissey, and G.~White, ``{Limits from BBN on Light Electromagnetic Decays},'' \href{https://dx.doi.org/10.1007/JHEP01(2019)074}{{\em JHEP} {\bfseries 01} (2019) 074}, \href{https://arxiv.org/abs/1809.01179}{{\ttfamily arXiv:1809.01179 [hep-ph]}}.

\bibitem{Hufnagel:2018bjp}
M.~Hufnagel, K.~Schmidt-Hoberg, and S.~Wild, ``{BBN constraints on MeV-scale dark sectors. Part II. Electromagnetic decays},'' \href{https://dx.doi.org/10.1088/1475-7516/2018/11/032}{{\em JCAP} {\bfseries 11} (2018) 032}, \href{https://arxiv.org/abs/1808.09324}{{\ttfamily arXiv:1808.09324 [hep-ph]}}.

\bibitem{Depta:2019lbe}
P.~F. Depta, M.~Hufnagel, K.~Schmidt-Hoberg, and S.~Wild, ``{BBN constraints on the annihilation of MeV-scale dark matter},'' \href{https://dx.doi.org/10.1088/1475-7516/2019/04/029}{{\em JCAP} {\bfseries 04} (2019) 029}, \href{https://arxiv.org/abs/1901.06944}{{\ttfamily arXiv:1901.06944 [hep-ph]}}.

\bibitem{Kawasaki:2017bqm}
M.~Kawasaki, K.~Kohri, T.~Moroi, and Y.~Takaesu, ``{Revisiting Big-Bang Nucleosynthesis Constraints on Long-Lived Decaying Particles},'' \href{https://dx.doi.org/10.1103/PhysRevD.97.023502}{{\em Phys. Rev. D} {\bfseries 97} no.~2, (2018) 023502}, \href{https://arxiv.org/abs/1709.01211}{{\ttfamily arXiv:1709.01211 [hep-ph]}}.

\bibitem{Poulin:2015opa}
V.~Poulin and P.~D. Serpico, ``{Nonuniversal BBN bounds on electromagnetically decaying particles},'' \href{https://dx.doi.org/10.1103/PhysRevD.91.103007}{{\em Phys. Rev. D} {\bfseries 91} no.~10, (2015) 103007}, \href{https://arxiv.org/abs/1503.04852}{{\ttfamily arXiv:1503.04852 [astro-ph.CO]}}.

\bibitem{Cyburt:2009pg}
R.~H. Cyburt, J.~Ellis, B.~D. Fields, F.~Luo, K.~A. Olive, and V.~C. Spanos, ``{Nucleosynthesis Constraints on a Massive Gravitino in Neutralino Dark Matter Scenarios},'' \href{https://dx.doi.org/10.1088/1475-7516/2009/10/021}{{\em JCAP} {\bfseries 10} (2009) 021}, \href{https://arxiv.org/abs/0907.5003}{{\ttfamily arXiv:0907.5003 [astro-ph.CO]}}.

\bibitem{Kusakabe:2008kf}
M.~Kusakabe, T.~Kajino, T.~Yoshida, T.~Shima, Y.~Nagai, and T.~Kii, ``{New Constraints on Radiative Decay of Long-Lived X Particles in Big Bang Nucleosynthesis with New Rates of Photodisintegration Reactions of $^4$He},'' \href{https://dx.doi.org/10.1103/PhysRevD.79.123513}{{\em Phys. Rev. D} {\bfseries 79} (2009) 123513}, \href{https://arxiv.org/abs/0806.4040}{{\ttfamily arXiv:0806.4040 [astro-ph]}}.

\bibitem{Jedamzik:2006xz}
K.~Jedamzik, ``{Big bang nucleosynthesis constraints on hadronically and electromagnetically decaying relic neutral particles},'' \href{https://dx.doi.org/10.1103/PhysRevD.74.103509}{{\em Phys. Rev. D} {\bfseries 74} (2006) 103509}, \href{https://arxiv.org/abs/hep-ph/0604251}{{\ttfamily arXiv:hep-ph/0604251}}.

\bibitem{Kohri:2005wn}
K.~Kohri, T.~Moroi, and A.~Yotsuyanagi, ``{Big-bang nucleosynthesis with unstable gravitino and upper bound on the reheating temperature},'' \href{https://dx.doi.org/10.1103/PhysRevD.73.123511}{{\em Phys. Rev. D} {\bfseries 73} (2006) 123511}, \href{https://arxiv.org/abs/hep-ph/0507245}{{\ttfamily arXiv:hep-ph/0507245}}.

\bibitem{Jedamzik:2004er}
K.~Jedamzik, ``{Did something decay, evaporate, or annihilate during Big Bang nucleosynthesis?},'' \href{https://dx.doi.org/10.1103/PhysRevD.70.063524}{{\em Phys. Rev. D} {\bfseries 70} (2004) 063524}, \href{https://arxiv.org/abs/astro-ph/0402344}{{\ttfamily arXiv:astro-ph/0402344}}.

\bibitem{Kawasaki:2004qu}
M.~Kawasaki, K.~Kohri, and T.~Moroi, ``{Big-Bang nucleosynthesis and hadronic decay of long-lived massive particles},'' \href{https://dx.doi.org/10.1103/PhysRevD.71.083502}{{\em Phys. Rev. D} {\bfseries 71} (2005) 083502}, \href{https://arxiv.org/abs/astro-ph/0408426}{{\ttfamily arXiv:astro-ph/0408426}}.

\bibitem{Kawasaki:2004yh}
M.~Kawasaki, K.~Kohri, and T.~Moroi, ``{Hadronic decay of late - decaying particles and Big-Bang Nucleosynthesis},'' \href{https://dx.doi.org/10.1016/j.physletb.2005.08.045}{{\em Phys. Lett. B} {\bfseries 625} (2005) 7--12}, \href{https://arxiv.org/abs/astro-ph/0402490}{{\ttfamily arXiv:astro-ph/0402490}}.

\bibitem{Cyburt:2002uv}
R.~H. Cyburt, J.~R. Ellis, B.~D. Fields, and K.~A. Olive, ``{Updated nucleosynthesis constraints on unstable relic particles},'' \href{https://dx.doi.org/10.1103/PhysRevD.67.103521}{{\em Phys. Rev. D} {\bfseries 67} (2003) 103521}, \href{https://arxiv.org/abs/astro-ph/0211258}{{\ttfamily arXiv:astro-ph/0211258}}.

\bibitem{Kohri:2001jx}
K.~Kohri, ``{Primordial nucleosynthesis and hadronic decay of a massive particle with a relatively short lifetime},'' \href{https://dx.doi.org/10.1103/PhysRevD.64.043515}{{\em Phys. Rev. D} {\bfseries 64} (2001) 043515}, \href{https://arxiv.org/abs/astro-ph/0103411}{{\ttfamily arXiv:astro-ph/0103411}}.

\bibitem{Kawasaki:2000qr}
M.~Kawasaki, K.~Kohri, and T.~Moroi, ``{Radiative decay of a massive particle and the nonthermal process in primordial nucleosynthesis},'' \href{https://dx.doi.org/10.1103/PhysRevD.63.103502}{{\em Phys. Rev. D} {\bfseries 63} (2001) 103502}, \href{https://arxiv.org/abs/hep-ph/0012279}{{\ttfamily arXiv:hep-ph/0012279}}.

\bibitem{Holtmann:1998gd}
E.~Holtmann, M.~Kawasaki, K.~Kohri, and T.~Moroi, ``{Radiative decay of a longlived particle and big bang nucleosynthesis},'' \href{https://dx.doi.org/10.1103/PhysRevD.60.023506}{{\em Phys. Rev. D} {\bfseries 60} (1999) 023506}, \href{https://arxiv.org/abs/hep-ph/9805405}{{\ttfamily arXiv:hep-ph/9805405}}.

\bibitem{Kawasaki:1994sc}
M.~Kawasaki and T.~Moroi, ``{Electromagnetic cascade in the early universe and its application to the big bang nucleosynthesis},'' \href{https://dx.doi.org/10.1086/176324}{{\em Astrophys. J.} {\bfseries 452} (1995) 506}, \href{https://arxiv.org/abs/astro-ph/9412055}{{\ttfamily arXiv:astro-ph/9412055}}.

\bibitem{Arbey:2018zfh}
A.~Arbey, J.~Auffinger, K.~P. Hickerson, and E.~S. Jenssen, ``{AlterBBN v2: A public code for calculating Big-Bang nucleosynthesis constraints in alternative cosmologies},'' \href{https://dx.doi.org/10.1016/j.cpc.2019.106982}{{\em Comput. Phys. Commun.} {\bfseries 248} (2020) 106982}, \href{https://arxiv.org/abs/1806.11095}{{\ttfamily arXiv:1806.11095 [astro-ph.CO]}}.

\bibitem{Pitrou:2019nub}
C.~Pitrou, A.~Coc, J.-P. Uzan, and E.~Vangioni, ``{Precision Big Bang Nucleosynthesis with the New Code PRIMAT},'' \href{https://dx.doi.org/10.7566/JPSCP.31.011034}{{\em JPS Conf. Proc.} {\bfseries 31} (2020) 011034}, \href{https://arxiv.org/abs/1909.12046}{{\ttfamily arXiv:1909.12046 [astro-ph.CO]}}.

\bibitem{Pisanti:2007hk}
O.~Pisanti, A.~Cirillo, S.~Esposito, F.~Iocco, G.~Mangano, G.~Miele, and P.~D. Serpico, ``{PArthENoPE: Public Algorithm Evaluating the Nucleosynthesis of Primordial Elements},'' \href{https://dx.doi.org/10.1016/j.cpc.2008.02.015}{{\em Comput. Phys. Commun.} {\bfseries 178} (2008) 956--971}, \href{https://arxiv.org/abs/0705.0290}{{\ttfamily arXiv:0705.0290 [astro-ph]}}.

\bibitem{Kawasaki:1994af}
M.~Kawasaki and T.~Moroi, ``{Gravitino production in the inflationary universe and the effects on big bang nucleosynthesis},'' \href{https://dx.doi.org/10.1143/PTP.93.879}{{\em Prog. Theor. Phys.} {\bfseries 93} (1995) 879--900}, \href{https://arxiv.org/abs/hep-ph/9403364}{{\ttfamily arXiv:hep-ph/9403364}}.

\bibitem{Carr:1974nx}
B.~J. Carr and S.~W. Hawking, ``{Black holes in the early Universe},'' \href{https://dx.doi.org/10.1093/mnras/168.2.399}{{\em Mon. Not. Roy. Astron. Soc.} {\bfseries 168} (1974) 399--415}.

\bibitem{Carr:1975qj}
B.~J. Carr, ``{The Primordial black hole mass spectrum},'' \href{https://dx.doi.org/10.1086/153853}{{\em Astrophys. J.} {\bfseries 201} (1975) 1--19}.

\bibitem{Conley:2006jg}
J.~A. Conley and T.~Wizansky, ``{Microscopic Primordial Black Holes and Extra Dimensions},'' \href{https://dx.doi.org/10.1103/PhysRevD.75.044006}{{\em Phys. Rev. D} {\bfseries 75} (2007) 044006}, \href{https://arxiv.org/abs/hep-ph/0611091}{{\ttfamily arXiv:hep-ph/0611091}}.

\bibitem{Argyres:1998qn}
P.~C. Argyres, S.~Dimopoulos, and J.~March-Russell, ``{Black holes and submillimeter dimensions},'' \href{https://dx.doi.org/10.1016/S0370-2693(98)01184-8}{{\em Phys. Lett. B} {\bfseries 441} (1998) 96--104}, \href{https://arxiv.org/abs/hep-th/9808138}{{\ttfamily arXiv:hep-th/9808138}}.

\bibitem{Myers:1986un}
R.~C. Myers and M.~J. Perry, ``{Black Holes in Higher Dimensional Space-Times},'' \href{https://dx.doi.org/10.1016/0003-4916(86)90186-7}{{\em Annals Phys.} {\bfseries 172} (1986) 304}.

\bibitem{Friedlander:2022ttk}
A.~Friedlander, K.~J. Mack, S.~Schon, N.~Song, and A.~C. Vincent, ``{Primordial black hole dark matter in the context of extra dimensions},'' \href{https://dx.doi.org/10.1103/PhysRevD.105.103508}{{\em Phys. Rev. D} {\bfseries 105} no.~10, (2022) 103508}, \href{https://arxiv.org/abs/2201.11761}{{\ttfamily arXiv:2201.11761 [hep-ph]}}.

\bibitem{Friedlander:2023qmc}
A.~Friedlander, N.~Song, and A.~C. Vincent, ``{Dark matter from higher-dimensional primordial black holes},'' \href{https://dx.doi.org/10.1103/PhysRevD.108.043523}{{\em Phys. Rev. D} {\bfseries 108} no.~4, (2023) 043523}, \href{https://arxiv.org/abs/2306.01520}{{\ttfamily arXiv:2306.01520 [hep-ph]}}.

\bibitem{Hooper:2019gtx}
D.~Hooper, G.~Krnjaic, and S.~D. McDermott, ``{Dark Radiation and Superheavy Dark Matter from Black Hole Domination},'' \href{https://dx.doi.org/10.1007/JHEP08(2019)001}{{\em JHEP} {\bfseries 08} (2019) 001}, \href{https://arxiv.org/abs/1905.01301}{{\ttfamily arXiv:1905.01301 [hep-ph]}}.

\bibitem{Hooper:2020evu}
D.~Hooper, G.~Krnjaic, J.~March-Russell, S.~D. McDermott, and R.~Petrossian-Byrne, ``{Hot Gravitons and Gravitational Waves From Kerr Black Holes in the Early Universe},'' \href{https://arxiv.org/abs/2004.00618}{{\ttfamily arXiv:2004.00618 [astro-ph.CO]}}.

\bibitem{Friedlander:2023jzw}
A.~Friedlander, N.~Song, and A.~C. Vincent, ``{Dark matter from hot big bang black holes},'' \href{https://dx.doi.org/10.1103/PhysRevD.108.L081301}{{\em Phys. Rev. D} {\bfseries 108} no.~8, (2023) L081301}, \href{https://arxiv.org/abs/2303.07372}{{\ttfamily arXiv:2303.07372 [hep-ph]}}.

\bibitem{Guedens:2002sd}
R.~Guedens, D.~Clancy, and A.~R. Liddle, ``{Primordial black holes in brane world cosmologies: Accretion after formation},'' \href{https://dx.doi.org/10.1103/PhysRevD.66.083509}{{\em Phys. Rev. D} {\bfseries 66} (2002) 083509}, \href{https://arxiv.org/abs/astro-ph/0208299}{{\ttfamily arXiv:astro-ph/0208299}}.

\bibitem{Majumdar:2002mra}
A.~S. Majumdar, ``{Domination of black hole accretion in brane cosmology},'' \href{https://dx.doi.org/10.1103/PhysRevLett.90.031303}{{\em Phys. Rev. Lett.} {\bfseries 90} (2003) 031303}, \href{https://arxiv.org/abs/astro-ph/0208048}{{\ttfamily arXiv:astro-ph/0208048}}.

\bibitem{Gonzalo:2022jac}
E.~Gonzalo, M.~Montero, G.~Obied, and C.~Vafa, ``{Dark dimension gravitons as dark matter},'' \href{https://dx.doi.org/10.1007/JHEP11(2023)109}{{\em JHEP} {\bfseries 11} (2023) 109}, \href{https://arxiv.org/abs/2209.09249}{{\ttfamily arXiv:2209.09249 [hep-ph]}}.

\bibitem{Montero:2022prj}
M.~Montero, C.~Vafa, and I.~Valenzuela, ``{The dark dimension and the Swampland},'' \href{https://dx.doi.org/10.1007/JHEP02(2023)022}{{\em JHEP} {\bfseries 02} (2023) 022}, \href{https://arxiv.org/abs/2205.12293}{{\ttfamily arXiv:2205.12293 [hep-th]}}.

\end{thebibliography}\endgroup

\end{document}